# Porous plates at incidence

Chandan Bose[1,2], Callum Bruce[1], Ignazio Maria Viola[1*]

[1] School of Engineering, Institute for Energy Systems, University of Edinburgh, Edinburgh, EH9 3FB, United Kingdom.
[2] Aerospace Engineering, School of Metallurgy and Materials, University of Birmingham, Birmingham, B15 2TT, United Kingdom.

*Corresponding author(s). E-mail(s): i.m.viola@ed.ac.uk;

**Abstract**

This paper investigates the effect of permeability on two-dimensional rectangular plates at incidences. The flow topology is investigated for Reynolds number ($Re$) values between 30 and 90, and the forces on the plate are discussed for $Re = 30$, where the wake is found to be steady for any value of the Darcy number ($Da$) and the flow incidence ($\alpha$). At $Re = 30$, for a plate normal to the stream and vanishing $Da$, the wake shows a vortex dipole with a stagnation point on the plate surface. With increasing $Da$, the separation between the vortex dipole and the plate increases; the vortex dipole shortens and is eventually annihilated at a critical $Da$. For any value of $Da$ below the critical one, the vortex dipole disappears with decreasing $\alpha$. However, at low $Da$, the two saddle-node pairs merge at the same $\alpha$, annihilating the dipole; while at high $Da$, they merge at different $\alpha$, resulting in a single recirculating region for intermediate incidences. The magnitudes of lift, drag, and torque decrease with $Da$. Nevertheless, there exists a range of $Da$ and $\alpha$, where the magnitude of the plate-wise force component increases with $Da$, driven by the shear on the plate's pressure side. Finally, the analysis of the fluid impulse suggests that the lift and drag reduction with $Da$ are associated with the weakening of the leading and trailing edge shear layer, respectively. The present findings will be directly beneficial in understanding the role of permeability on small permeable bodies.

## 1 Introduction

The flow around porous bodies has a wide range of applications, including aerospace [1–4] and environmental [5, 6] engineering, and offshore energy [7]. Most of these applications are concerned with high Reynolds number ($Re$) flows, where the wake of the permeable body is turbulent. However, some biological flyers, such as bristled-wing insects [8] and plant seeds [9], and applications such as bio-inspired man-made microflyers for environmental monitoring [10] are concerned with low *Re* flows. The main challenge of efficiently designing these devices is to minimise the weight to increase the time afloat. In this regard, several studies have exploited permeability to decrease weight, including using filamentous structures and porous materials [10–12]. The filament-based *Re* of these structures is in the order of 1 to 100. However, the effect of permeability on the aerodynamics of flow-immersed bodies at low *Re* is yet to be fully understood. To this end, this paper considers the effect of permeability for a cylinder with an infinite span and a rectangular cross-section with the width-to-thickness ratio ($\chi$) of 10, immersed in a uniform and constant flow stream at *Re* between 30 and 90.

The flow field past a cylinder with a rectangular cross-section, such as in the present study, as well as with a circular cross-section, is two-dimensional (2D) within this *Re* range considered in this work [13–16]. Specifically, the wake remains 2D up to a critical *Re* around 200 for impervious cylinders in cross flow [17, 18]. Therefore, in this work, we consider the flow on a 2D section of a cylinder with an infinite span and a rectangular cross-section, a 2D plate hereafter.

The flow past 2D *impervious* plates has been extensively studied experimentally [19–24], numerically [15, 20, 21, 25–29], as well as theoretically [30]. When the plate is orthogonal to the stream, i.e. at an incidence $\alpha = 90°$, flow separation occurs at the edges of the plate [30]. For Reynolds number (*Re*) values below a critical threshold, which decreases with $\chi$, a steady vortex dipole is formed in the wake of the plate [31]. The critical *Re* decreases, for example, from 115 to 30 for $\chi = 1/4$ and 10, respectively [30, 32]. The vortex dipole is attached to the plate with its upstream stagnation point on the plate surface. The downstream stagnation point shifts downstream with increasing *Re*, resulting in a greater vortex dipole, and vortex shedding occurs for *Re* higher than the critical threshold [15].

How the flow topology varies with $\alpha$ is known for an *impervious* plate. At $\alpha = 0$, the plate is aligned with the flow, but separation occurs due to the blunt leading edge if $\chi > 0$. For $\chi$ greater than a threshold that depends on *Re*, the flow reattaches forming a leading edge bubble [33, 34], and the flow remains attached up to the downstream blunt trailing edge. If *Re* is sufficiently low to ensure a steady wake, increasing $\alpha$, the flow topology changes from attached flow to a single vortex, and eventually to a vortex dipole as $\alpha$ approaches 90° [26].

The effect of porosity on the flow past a 2D plate has received considerable attention because of its wide applications at high *Re*, such as grids, air brakes, parachutes, wind and water turbines, etc. These studies reveal that the governing parameters are *Re*, the nondimensional permeability expressed by the Darcy number *Da*, and the porosity $\varepsilon$ [35, 36]. While *Da* and $\varepsilon$ are independent parameters, for a constant *Da*, $\varepsilon$ only weakly affects the flow field and its stability [37]. Betz [38] was among the first to develop a theoretical model for the drag of a permeable plate placed orthogonal to the flow. The model is based on potential flow away from the plate and a momentum budget around the plate. The model has been further developed by Taylor [39], who used a distribution of sources to account for the wake, and several other authors [e.g. 40], including the recent work of Steiros and Hultmark [41], which adds the drag due to the base suction. However, these works apply to sufficiently high *Re*, where viscous forces can be neglected or confined to the permeable region, and focus on the effect of $\varepsilon$ rather than *Da*.

Furthermore, the effect of the inclination of the plate with respect to the freestream is not considered in these studies. To this end, Woods [42], and more recently also Hajian and Jaworski [43] considered a permeable airfoil in an inviscid flow, where viscosity is confined in the boundary layer and the permeable region in which a Darcy porosity condition is applied. Hence, these studies also do not apply to low-*Re* flows where the viscous effects are significant and only provide pressure forces from the derivation of the pressure distribution around the foil.

The flow field around a permeable flat plate, and bluff body in general, can be inferred from several numerical and experimental studies on porous squared cylinders [37, 44–46], circular cylinders [47], spheres [48, 49], and axisymmetric disks [37, 50–52]. In general, in the wake of permeable bluff bodies, a steady recirculation region exists for low *Re* and low *Da*; vortex shedding occurs for high *Re* and low *Da*; while the wake is steady without recirculation for high *Da*. The *Re* and *Da* thresholds depend on the geometry [37, 52]. For example, for a permeable disk orthogonal to the flow and with $\chi = 10$, the maximum *Da* at which a recirculation region exists is $10^{-3}$ with a mild non-monotonic dependency on *Re* [50, 52].

Two-dimensional permeable plates have only been investigated experimentally [e.g. 53, 54] and numerically [55] at relatively high *Re* and at $\alpha = 90^\circ$, i.e. where a vortex dipole exists only in the time average sense for low *Da*. Instead, the effect of permeability on 2D plates at low *Re*, where the wake is steady, and various $\alpha$ and the associate wake topologies have not been investigated before. To that end, in this paper, we numerically study the flow past 2D porous plates with $\chi = 10$ and porosity $\varepsilon = 0.95$, for a range of *Da*, and $\alpha$ values. We first study the steady-to-unsteady wake transition for a range of *Re* values, and then carry out a detailed flow-field and force analysis at $Re = 30$, where the wake remains steady throughout the chosen parametric space.

The primary objectives of this paper are the following: i) to identify the *Re*, *Da*, and $\alpha$ range where the wake of a permeable plate is steady; ii) to identify how the aerodynamic loads and the topology of the steady wake varies with *Re*, *Da*, and $\alpha$; and iii) to reveal, through the application of the impulse theory, how the vorticity field is correlated with the aerodynamic loads.

The remainder of the paper is structured as follows. The numerical method, including detailed domain size and grid resolution independent study and solver validation, is presented in §2. The results and discussions are presented in §3. These include, first, the identification of the transition from a steady to an unsteady wake in the $Re - Da$ parameter space (§3.1); then the analysis of the flow-field past a plate normal to the stream (§3.2) and at different incidences with the stream (§3.3); finally, the analysis of how the forces and torque change with $\alpha$ and *Da* (§3.4) and the identification, using impulse theory, of the associated changes in the vorticity field (§3.5). The salient outcomes of this study are summarised in §4.

## 2 Methodology

We model a 2D porous plate with width $\hat{d}$ and thickness $\hat{f}$ in a uniform stream of fluid with density $\hat{\rho}$ and velocity $\hat{\boldsymbol{u}}_\infty$. The hat over the symbols is used to indicate dimensional quantities. In the following, all quantities are made nondimensional using the base $(\hat{\rho}, \hat{d}, \hat{\boldsymbol{u}}_\infty)$. We define two different frames of reference (Fig. 1a): (1) a global frame of reference $\mathrm{O}(X,Y)$, where $X$ and $Y$ are parallel and orthogonal to $\hat{\boldsymbol{u}}_\infty$, respectively; and (2) a body fixed frame of reference $\mathrm{O}(x,y)$, where $x$ and $y$ are in the plate-normal and plate-wise direction, respectively. The angle of attack $\alpha$ is defined as the complementary angle to the angle between the two frames of references, and thus $O(X,Y) = O(x,y)$ when $\alpha = 90°$. Figure 1b schematically presents the separated vortex dipole in the wake of a porous plate normal to the stream, where the saddle points and nodes are marked by S1, S2, and N1, N2, respectively.

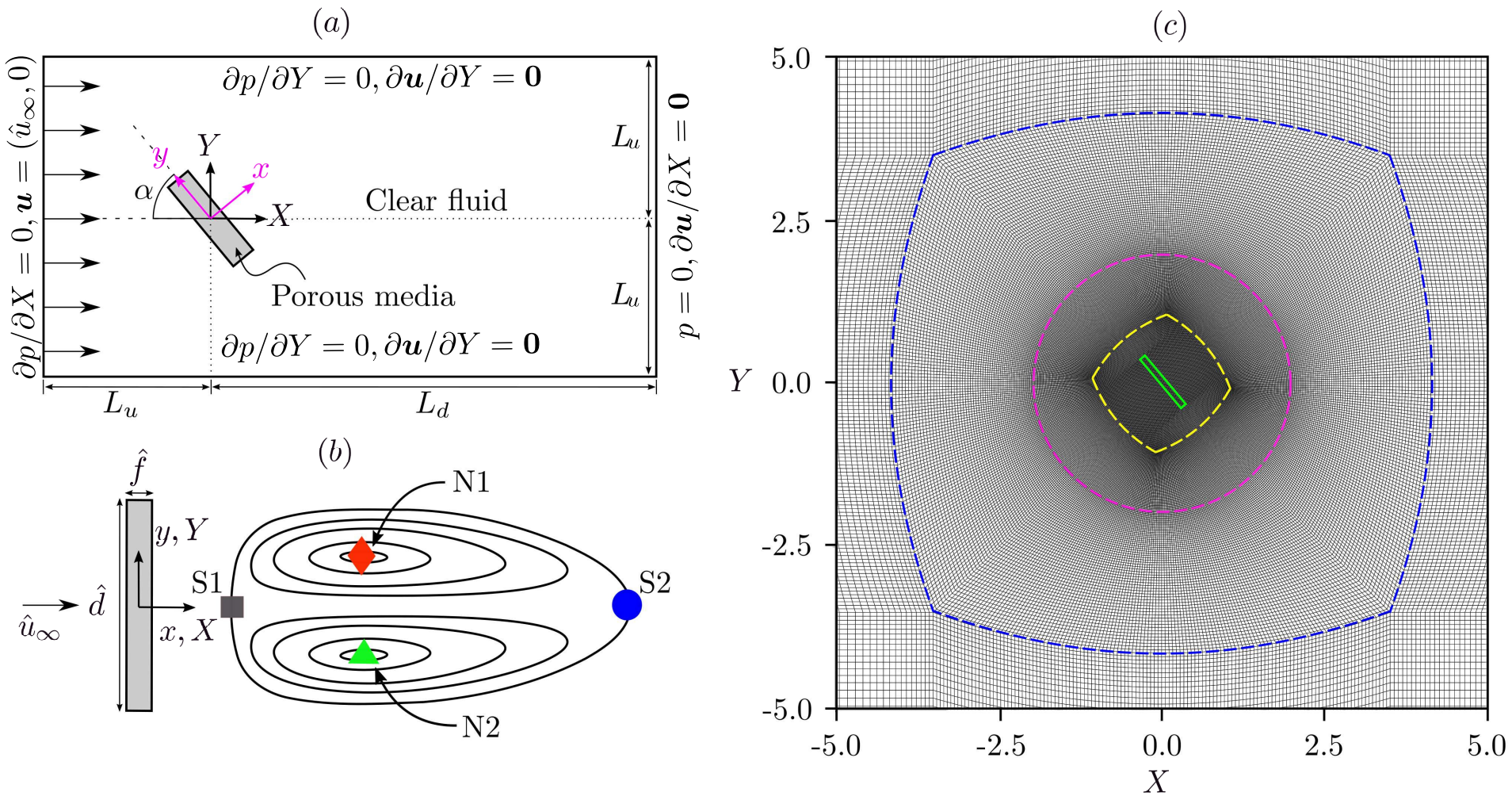

**Fig. 1**: (a) Computational domain and boundary conditions (not to scale). (b) Schematic of the vortex dipole in the wake of a porous plate normal to the stream; saddle points S1, S2, and nodes N1, N2 are labelled. (c) Computational grid near the porous plate (green) at $\alpha = 50°$. The external H-type grid transforms to O-type on the blue dashed line, and back to H-type on the yellow dashed line. The magenta dashed line is the interface about which the internal part of the grid is rotated for varying $\alpha$.

## 2.1 Governing equations and numerical approach

We solve, in the $O(X,Y)$ frame, the continuity equation and the Darcy-Brinkman-Forchheimer equation [35, 36, 56]. The governing equations, in nondimensional form, are

$$\nabla \cdot \boldsymbol{u} = 0, \tag{1}$$

$$\frac{1}{\varepsilon}\frac{\partial \boldsymbol{u}}{\partial t} + \frac{1}{\varepsilon^2}(\boldsymbol{u}\cdot\nabla)\boldsymbol{u} = -\nabla p + \frac{1}{\varepsilon Re}\nabla^2\boldsymbol{u} - \frac{1}{ReDa}\boldsymbol{u} - \frac{c_F}{\sqrt{Da}}|\boldsymbol{u}|\boldsymbol{u}, \tag{2}$$

where $\boldsymbol{u} = (u,v)$ is the nondimensional velocity vector with components $u$ and $v$; $t$ is the nondimensional time; $p$ is the nondimensional pressure; $Re = \hat{u_\infty}\hat{d}/\hat{\nu}$ is the Reynolds number, $\hat{\nu}$ is the kinematic viscosity, $Da = \hat{K}/\hat{d}^2$ is the Darcy number, $\hat{K}$ is the dimensional permeability, $\varepsilon$ is the porosity, which is defined as the ratio between the empty area of pores and the total area of the chord-normal projection of the plate; and $c_F$ is a form-drag coefficient.

We solve Eq. 2 for both the clear fluid and the porous regions, each with different values of Darcy number ($Da$) and porosity ($\varepsilon$). In fact, the governing equation for the clear fluid region is the incompressible Navier-Stokes equation, which is the asymptotic form of the Darcy-Brinkman-Forchheimer equation (Eq. 2) for $Da \to \infty$ and $\varepsilon = 1$. In our simulations, we use a high value of $Da = 10^{12}$ and $\varepsilon = 1$ for the clear fluid region.

In this study, we consider $\varepsilon = 0.95$ and $c_F = 0$, but different values are adopted to validate our results with those of other authors (see Sec. 2.4). We have developed a customised porous incompressible Navier-Stokes solver, `porousIcoFoam`, by modifying the `icoFoam` solver, available within the finite volume method based open-source library `OpenFOAM`. The spatial and temporal discretisations are second-order accurate. The Pressure Implicit with the Splitting of the Operator algorithm with a predictor step and two pressure correction loops has been used to couple the pressure and velocity equations. A preconditioned conjugate gradient iterative solver is used to solve the pressure equation, whereas a diagonal incomplete-Cholesky method is used for preconditioning. A preconditioned smooth solver is used to solve the pressure-velocity coupling equation, and the symmetric Gauss-Seidel method is used for preconditioning. The absolute error tolerance criteria for pressure and velocity are set to $10^{-6}$. The simulations are run for 240 convective periods ($0 \leqslant t \leqslant 240$). This ensures convergence of the steady-state loads with relative errors smaller than $10^{-8}$. The simulations took around three hours with 128 processors in the Archer2 HPC system (Edinburgh, UK).

## 2.2 Computational domain and mesh

The porous plate is placed within a rectangular computational domain with the edges parallel to the $X$ and $Y$ axes. The plate is placed at a distance $L_u = 20$ from the upstream and the lateral sides of the domain and $L_d = 80$ from the downstream edge (Fig. 1a). At the upstream edge (inlet), we set $\boldsymbol{u} = (u_\infty, 0)$ and $\partial p/\partial X = 0$, while at the downstream edge (outlet), $p = 0$ and $\partial \boldsymbol{u}/\partial X = \mathbf{0}$. On the side edges, we apply a slip condition with $\partial p/\partial Y = 0$ and $\partial \boldsymbol{u}/\partial Y = \mathbf{0}$. At the edges of the plate, which is the interface between the clear fluid and the porous media, velocity, pressure, and stresses are conserved. The domain is made of an external region, which is fixed for all simulations, and an internal region, which is rotated by $\alpha$. The grid topology is shown in Fig. 1c. The initial condition is a uniform flow on the whole domain with $\boldsymbol{u} = (u_\infty, 0)$ and $p = 0$.

Motivated by the requirement that $\alpha$ should be varied while maintaining grid topology and connectivity, a structured H-O-H type hexahedral grid is employed. A radial coordinate, $r$ is defined from the origin as $r = \sqrt{x^2 + y^2}$ where $x$ and $y$ are the local Cartesian coordinates. Starting from the outermost portion of the grid and moving in towards the origin, the grid is initially of the H-type. The grid changes from H-type to O-type and transforms from a square to circular topology, between $r \sim 3.5$ and $r = 2$. The grid changes back from O-type to H-type, transforming from a circular to square topology between $r = 2$ and $r \sim 0.75$. The porous flat plate is contained within the inner H-type portion of the grid. Only cells in the region $r < 2$ are transformed when $\alpha$ is varied. Grid connectivity is maintained along the boundary where $r = 2$. Grid spacing at the external boundaries of the domain is described by the number of cells along the inlet and outlet patches, $n_y$, and the number of cells along the upper and lower boundaries of the domain, $n_x$. The numbers of grid points along the two external domain boundaries are 492 and 295, respectively, and 84 along the plate width and 8 along the plate thickness. Uniform grid spacing is maintained inside the porous region.

## 2.3 Forces calculations

The fluid forces generated by the plate are computed from the pressure and shear forces acting on the edges of the plate, i.e. at the interface between the clear fluid and the porous media. Once the steady state is reached, the force is

$$\boldsymbol{F} = \oint_l p\boldsymbol{n}\,\mathrm{d}l + \oint_l \mathscr{S}\cdot\boldsymbol{n}\,\mathrm{d}l, \tag{3}$$

where $\boldsymbol{n}$ is the unit vector locally normal to the plate perimeter $l$ and pointing outwards; $\mathscr{S}$ is the viscous stress tensor. It is noted that using the base $(\hat{\rho}, \hat{d}, \hat{u_\infty})$, the force and torque coefficients are twice the nondimensional forces and torque. The lift, drag, and torque coefficients are $C_L = 2L, C_D = 2D$, and $C_M = 2M$, respectively. The torque is computed with respect to the origin of the frames and is positive anticlockwise.

## 2.4 Verification and validation

To estimate the modelling error due to the finite dimension of the domain, and the numerical error due to the finite cell size, we consider three cases and compare the results with those of other authors. Case 1 is a steady simulation of the flow past a rectangular cylinder (i.e. $\chi = 1$) with $Re = 30$, $Da = 10^{-3}$, $\varepsilon = 0.977$, and $c_F = 0.148$. This case was modelled by Dhinakaran and Ponmozhi [44] and by Anirudh and Dhinakaran [45]. Case 2 is an unsteady simulation of the same geometry ($\chi = 1$) with $Re = 75$, $Da = 10^{-6}$, $\varepsilon = 0.629$, and $c_F = 0.286$. Finally, Case 3 is that modelled by Ledda et al. [37]: a slender plate normal to the flow with $\chi = 10$, where $Re = 30$, $Da = 1.1 \times 10^{-3}$, $\varepsilon = 0.650$, and $c_F = 0$. For these three cases, we consider the errors in the estimates of $C_D$, the $X$-coordinate of the downstream saddle point S2, denoted by $X_{\mathrm{S2}}$, and the Strouhal number $St$. We follow the verification and validation procedure outlined in Viola et al. [57]. This method was originally developed for yacht sails and was successively adopted for a wide range of applications, including the flow past the pappus of the dandelion diaspore [58], permeable disks [50], oscillating flapping foils [59], wind turbines [60], tidal turbines [61], arrays of energy harvesters [62], and ship hulls [63].

We consider four domains and three grids. The domains are built by progressively extending $L_u$ and $L_d$ by steps of 5 and 20, respectively. Specifically, $L_u = 10, 15, 20, 25$ and $L_d = 40, 60, 80, 100$ for domain D1 to D4, respectively. The grid spacing is the same for all domains, while the total number of grid points increases from D1 to D4. The G2 and G3 grids are achieved by scaling each cell size of G1 by $\sqrt{2}$ and 2, respectively. Hence, the number of grid points along the domain boundaries is $n_X = 350, 492, 700; n_Y = 210, 295, 420$; and along the width and thickness of the plate is $n_d = n_t = 60, 84, 120$ for G1 to G3, respectively. The base grid G2 is used for the domain size investigation, and the base domain D3 is used for the grid resolution investigation.

We consider the relative change ($\phi$) of a generic scalar with respect to the value computed with the base setting, with the relative change ($h$) of a source of error. The latter is chosen such that $h \to 0$ when the source of error vanishes. For example, Figure 2a shows the relative change of the drag coefficient, $\phi_{C_D} = C_D/C_{D_{\mathrm{base}}}$, with the inverse of the relative domain size, $h = h_d = (L_u/L_{u_{\mathrm{base}}})^{-1} = 20/L_u$. Figure 2b shows the relative change of S2's $X$-coordinate, $\phi_{X_{\mathrm{S2}}} = X_{\mathrm{S2}}/X_{\mathrm{S2}_{\mathrm{base}}}$, with the inverse of the relative number of cells, $h = h_g = (n_X/n_{X_{\mathrm{base}}})^{-1} = 492/n_X$.

We fit the data with $\phi = ch^p + \phi_0$, where the coefficients $c$, $p$ and $\phi_0$ are computed with least square optimisation, and the standard deviation of the residuals is $\sigma$. The advantage of presenting the data in this form is that the extrapolated value $\phi_0$ for $h \to 0$ is the expected true value of $\phi$. For example, in Figure 2a and 2b, the extrapolated values $\phi_0$ are ca. 0.96 and 1.005, which is about 4% lower and 0.05% higher than those computed with the base domain and grid. Hence, these latter values computed as $\delta = 1 - \phi_0$, are the estimated errors. This procedure is used for both the modelling error due to the domain size, and the numerical error due to the grid size. Table 1 shows the modelling errors $\delta_{C_D}$, $\delta_{X_{S2}}$, and $\delta_{St}$ in the computation of $C_D, X_{S2}$ and $St$, respectively, for Case 1 and Case 2.

For the numerical error due to the grid resolution, we compute the 95% confidence interval $[-U_\phi, U_\phi]$ centred on the value computed with the base setting (G2). The uncertainty $U_\phi$ is computed differently depending on the order of convergence $p$ of the least square fit. Specifically, for $p \geqslant 0.95$, $U_\phi = 1.25|\delta_\phi| + \sigma$. For $p < 0.95$, $U_\phi = 1.5\Delta_\phi + \sigma$, where $\Delta_\phi = [\max(\phi) - \min(\phi)]/[1 - \min(h)/\max(h)]$. This estimate is valid for any $p < 0.95$, but when $-0.05 \leqslant p \leqslant 0.05$, the confidence interval can alternatively be centred on the mean of all the computed values, and the uncertainty estimated as $U_{\phi_{mean}} = 2(\sigma_\phi/\sqrt{N})$, where $N$ is the number of step sizes used and $\sigma_\phi$ is the standard deviation of the distribution of $\phi$. Here, we adopt this second approach for $St$. Table 2 shows the uncertainty in the computation of $C_D, X_{S2}$ and $St$ for Case 1 and Case 2. The values of the $p$ coefficient for the cases shown in table 2 are as follows. Case 1: $p_{C_D} = 0.57$, $p_{X_{S2}} = 1.56$. Case 2: $p_{C_D} = 0.02$, $p_{St} = 0.05$.

**Table 1**: Modelling error due to the finite domain size for $C_D$, $X_{S2}$ and $St$ for a porous square cylinder. Case 1: $\chi = 1, Re = 30, Da = 10^{-3}, \varepsilon = 0.977, c_F = 0.148$ (steady); Case 2: $\chi = 1, Re = 75, Da = 10^{-6}, \varepsilon = 0.629, c_F = 0.286$ (unsteady).

| | | Case 1 | | | | Case 2 | | | |
|---|---|---|---|---|---|---|---|---|---|
| Domain | $h_d$ | $C_D$ | $\delta_{C_D}(\%)$ | $X_{S2}$ | $\delta_{X_{S2}}(\%)$ | $C_D$ | $\delta_{C_D}(\%)$ | $St$ | $\delta_{St}(\%)$ |
| D1 | 0.50 | 1.9444 | 8.76 | 1.9455 | 1.01 | 1.5244 | 6.14 | 0.1366 | 4.92 |
| D2 | 0.75 | 1.8763 | 5.07 | 1.9431 | 0.89 | 1.4834 | 3.34 | 0.1337 | 2.73 |
| D3 | 1.00 | 1.8463 | 3.45 | 1.9371 | 0.58 | 1.4664 | 2.18 | 0.1324 | 1.75 |
| D4 | 1.25 | 1.8297 | 2.55 | 1.9321 | 0.32 | 1.4572 | 1.55 | 0.1318 | 1.30 |

**Table 2**: Uncertainty due to finite grid resolution for $C_D$, $X_{S2}$ and $St$ for a porous square cylinder. Case 1: $\chi = 1, Re = 30, Da = 10^{-3}, \varepsilon = 0.977, c_F = 0.148$ (steady); Case 2: $\chi = 1, Re = 75, Da = 10^{-6}, \varepsilon = 0.629, c_F = 0.286$ (unsteady).

| | | Case 1 | | Case 2 | |
|---|---|---|---|---|---|
| Grid | $h_g$ | $C_D, U_{C_D}(\%)$ | $X_{S2}, U_{X_{S2}}(\%)$ | $C_D, U_{C_D}(\%)$ | $St, U_{St}(\%)$ |
| G1 | $\sqrt{2}$ | 1.8306, 4.65 | 1.9292, 1.22 | 1.4637, 0.21 | 0.1331, 1.27 |
| G2 | 1 | 1.8463, 4.65 | 1.9371, 0.71 | 1.4664, 0.21 | 0.1324, 1.27 |
| G3 | $1/\sqrt{2}$ | 1.8592, 4.65 | 1.9417, 0.41 | 1.4609, 0.21 | 0.1303, 1.27 |

The values of $C_D, X_{S2}$ and $St$ computed with the base setting (D3, G2) are compared with those of other authors in table 3. The differences are consistent with the modelling errors due to the finite domain size and the numerical uncertainty. Note that Dhinakaran and Ponmozhi [44] and Anirudh and Dhinakaran [45] have used domains that are about half of D3 in size, and this is consistent with their higher estimates of $C_D, X_{S2}$ and $St$.

Overall, the numerical and modelling error analysis reveals that the forces are computed within a numerical uncertainty at a 95% confidence level of 4.65% of the computed value, while the coordinates of topological points in the wake are computed within an uncertainty of 0.71% of the computed value, or $1.9371 \times 0.71\% = 1.37\%$ of the chord length. The error due to the finite size of the domain is estimated at 5.07% and 0.89% of the computed values, respectively.

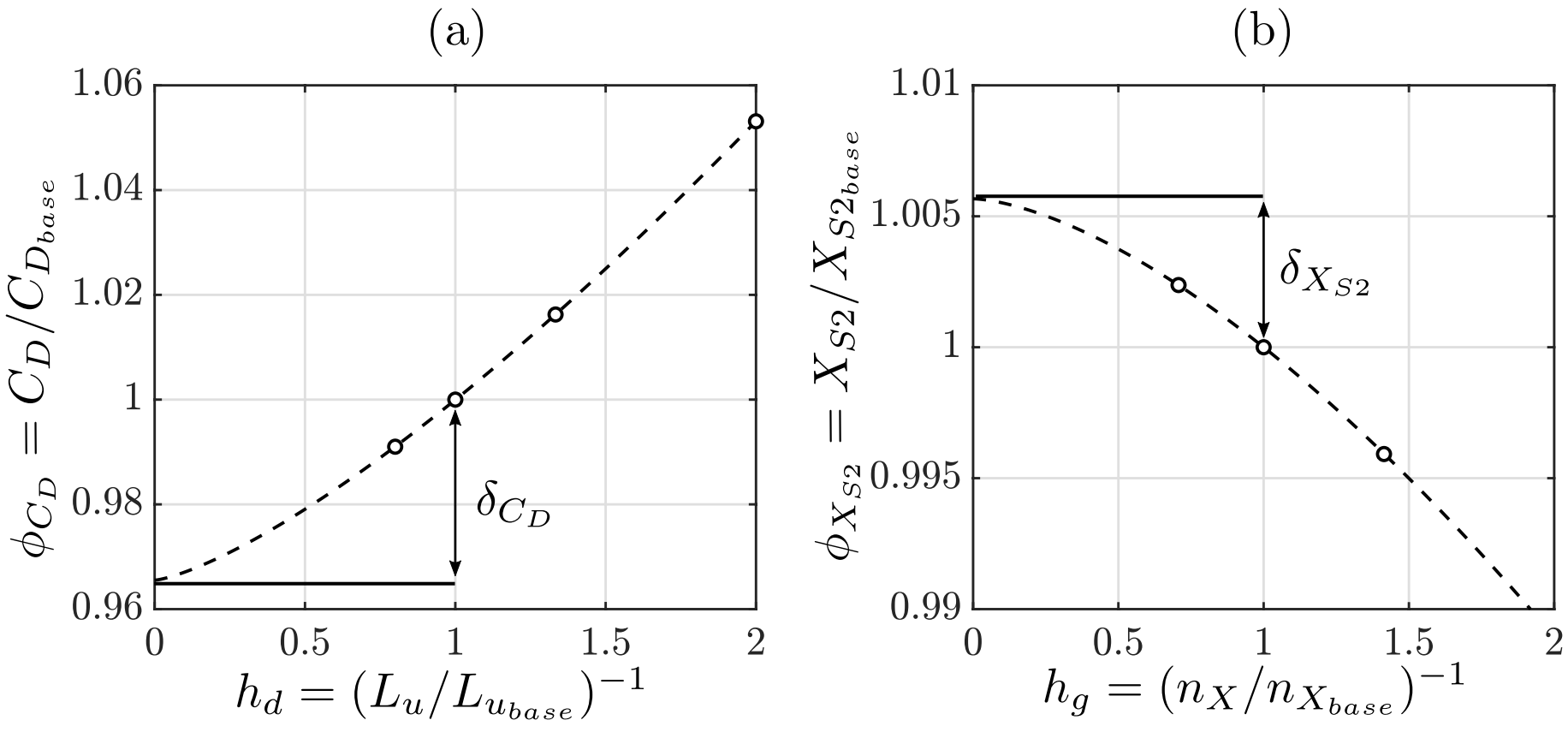

**Fig. 2**: Convergence of the drag coefficient with the domain size (a) and of the $X$-coordinate of the downstream saddle point with the grid resolution (b) for Case I.

**Table 3**: Present study results compared with three cases from the literature. Case 1: $\chi = 1, Re = 30, Da = 10^{-3}, \varepsilon = 0.977, c_F = 0.148$ (steady); Case 2: $\chi = 1, Re = 75, Da = 10^{-6}, \varepsilon = 0.629, c_F = 0.286$ (unsteady); Case 3: $\chi = 4, Re = 30, Da = 1.1 \times 10^{-3}, \varepsilon = 0.650, c_F = 0.0$.

| | Case 1 | | Case 2 | | Case 3 | |
|---|---|---|---|---|---|---|
| | $C_D$ | $X_{S2}$ | $C_D$ | $St$ | $C_D$ | $X_{S2}$ |
| Anirudh and Dhinakaran (2018) | 2.0021 | 1.9820 | 1.5639 | 0.1374 | - | - |
| Dhinakaran and Ponmozhi (2011) | 2.0255 | 1.9415 | - | - | - | - |
| Sharma and Eswaran (2004) | - | - | 1.5490 | 0.1370 | - | - |
| Ledda et al. (2018) | - | - | - | - | 1.86 | 2.15 |
| Present study | 1.8463 | 1.9371 | 1.4664 | 0.1324 | 1.77 | 2.07 |

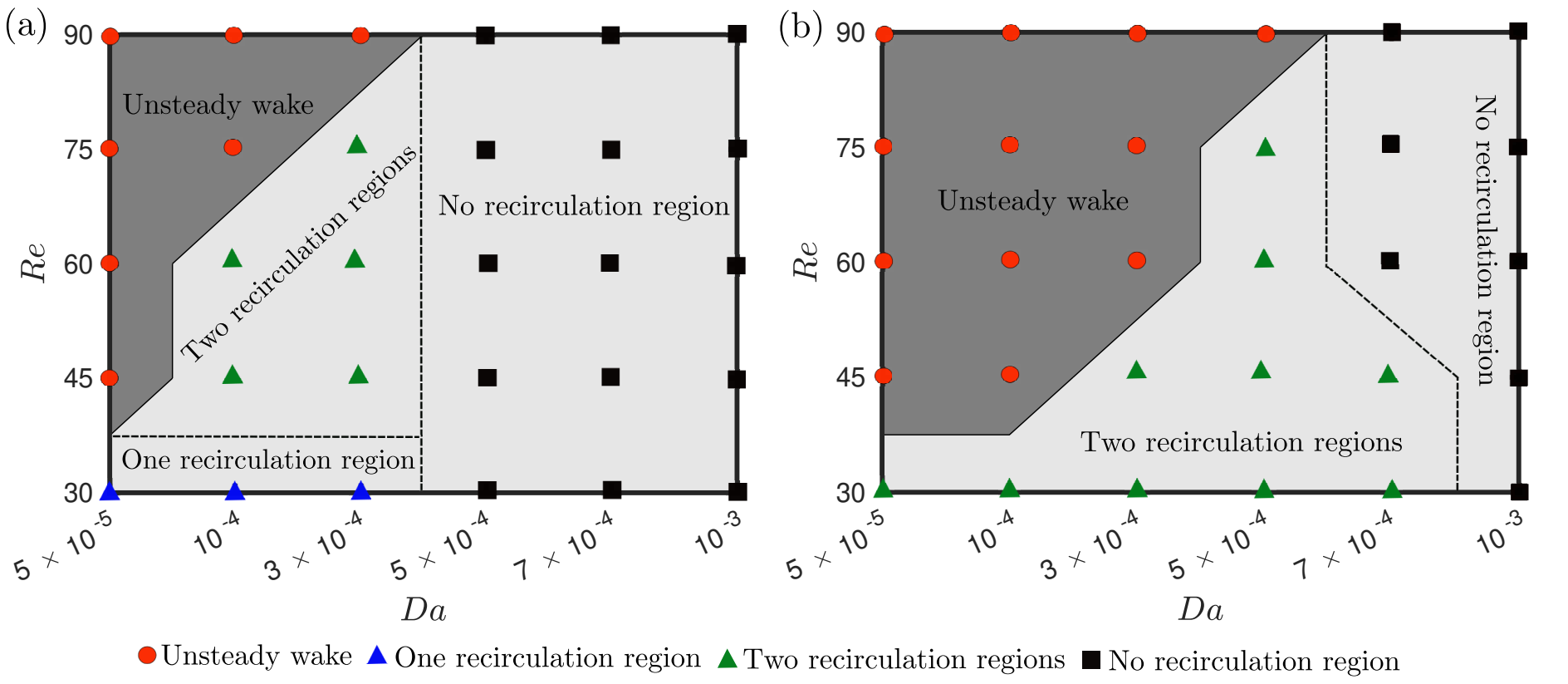

**Fig. 3**: Map of the wake topology for a 2D porous plate for a range of Reynolds and Darcy numbers: (a) $\alpha = 40°$ and (b) $\alpha = 90°$. The light grey region and the deep grey region represent the steady and unsteady wake regions, respectively. The red circles, blue and green triangles, and black squares are indicative of unsteady vortex shedding, one and two recirculation regions, and no recirculation, respectively.

# 3 Results and Discussions

## 3.1 Steady-unsteady transition boundary

First, we investigate the *Re* and *Da* values for which the wake is steady and unsteady for flow incidences $\alpha = 40°$ and $90°$ (figs. 3a and 3b, respectively). At both incidences, the critical *Re* value at which steady-unsteady transition occurs increases with *Da*, and decreases with $\alpha$ (solid line in figs. 3a and 3b). We also identify the transition between a steady wake with and without a recirculation region (dashed line in figs. 3a and 3b). Viscous forces dominate at low *Re*, and the porous medium imposes minimal resistance at high *Da* values, resulting in a steady wake. The shear layers formed at the edges of the plate become thinner with increasing *Re* and stronger with decreasing *Da*, gradually leading to shear layer instabilities that grow into unsteady transition.

A reduction in the angle of incidence from 90° has a stabilizing effect, shifting the steady-to-unsteady boundary towards higher *Re* and lower *Da*. This can be associated with a reduction in the effective length scale in the cross-flow direction (that is $\hat{d}\cos\alpha$), in turn lowering the effective *Re*. Hence, the wake remains steady for higher *Re* and lower *Da* values at, for example, $\alpha = 40°$ compared to $\alpha = 90°$.

The loss of longitudinal symmetry at $\alpha < 90°$ allows new topology to emerge, including a steady single recirculation region at $\alpha = 40°$. The transformation of the wake from two symmetric recirculation regions at $\alpha = 90°$ to a single recirculation region at $\alpha = 40°$ is associated with the gradual convergence and then merging of the nodes and the saddle points, as discussed in Sec. 3.3.

Finally, the "no recirculation region" boundary is shifted towards lower *Da* for $\alpha = 40°$ compared to $\alpha = 90°$. This can be associated with the lower drag and pressure drop across the permeable plate, and the higher base pressure on the leeward side of the plate. Because the difference between the base pressure and the free stream pressure is lower for $\alpha = 40^\circ$ than for $\alpha = 90^\circ$, the flow does not need to trade as much velocity to increase the pressure to the free stream level along the wake. In the rest of the paper, we investigate how the permeability allows switching between these three different flow topologies, and we focus on $Re = 30$ to ensure that the wake remains steady.

## 3.2 Porous plate normal to the stream

We first consider a plate at $Re = 30$ and $\alpha = 90^\circ$, where two recirculation regions are formed as shown in Fig.1b, and we investigate the effect of permeability. Figure 4a shows the $X$ coordinate of the upstream (S1) and downstream (S2) saddle points versus $Da$, while Fig. 4b shows $C_D$ versus $Da$. For vanishing permeability ($Da \to 0$), the flow topology and the force tend asymptotically towards those of an impervious plate. The small differences between the results for an impervious plate and the asymptotic values for vanishing $Da$ are attributed to the different numerical algorithms for porous and impervious bodies. As $Da$ increases, S1 moves downstream, and S2 moves upstream, shrinking the vortex dipole up to its annihilation at a critical $Da$, between $8 \times 10^{-4}$ and $9 \times 10^{-4}$. This is also shown in figs. 4c-4e by means of the streamlines superimposed to the vorticity contour. The reduction in size and eventual annihilation of the vortex dipole coincides with a reduction in $C_D$. The $C_D$ values obtained for an axisymmetric porous disc by Cummins et al. [50] at $Re = 30$ are relatively higher than those obtained for a 2D flat plate in this study. However, the trend of $C_D$ variation with respect to $Da$ is qualitatively similar in these two cases.

## 3.3 Flow around a porous plate at incidence

We now turn our study to the effect of the angle of incidence on permeable plates. With decreasing $\alpha$ from $90^\circ$ to $0^\circ$, the vortex dipole first becomes asymmetric and eventually annihilates. This occurs through different topological steps for low and high permeability values. For plates with low permeability, as well as for the impervious plates, the recirculating wake annihilates in two steps as $\alpha$ is decreased. Conversely, beyond a critical $Da$ value, between $5 \times 10^{-5}$ and $5 \times 10^{-4}$, the vortex dipole annihilates in a single step. This is shown in figs. 5a and 5b for low and high permeability cases ($Da = 5 \times 10^{-5}$ and $5 \times 10^{-4}$), respectively, where the body-fixed coordinates of the topological points (N1, N2, S1, and S2) are tracked for various incidences.

For the low-permeability case, with decreasing $\alpha$, all four topological points move first downstream parallel with $y$ and then turn towards the plate (Fig. 5c). At $\alpha = 44^\circ$, N2 merges with S2, forming a distinct topological field, comprising only one re-circulation region with closed streamlines and negative circulation (Fig. 5d). This topology exists for $\alpha$ as low as $30^\circ$, when N1 merges with S1 annihilating the re-circulation region (Fig. 5e). In contrast, for the high-permeability case, the intermediate topology with only one re-circulation region does not exist (Fig. 5f). Instead, both node-saddle point pairs merge at $\alpha = 64^\circ$ (Fig. 5g). For lower values of $\alpha$, the wake is characterised by tortuous streamlines with no closed re-circulation regions (Fig. 5h).

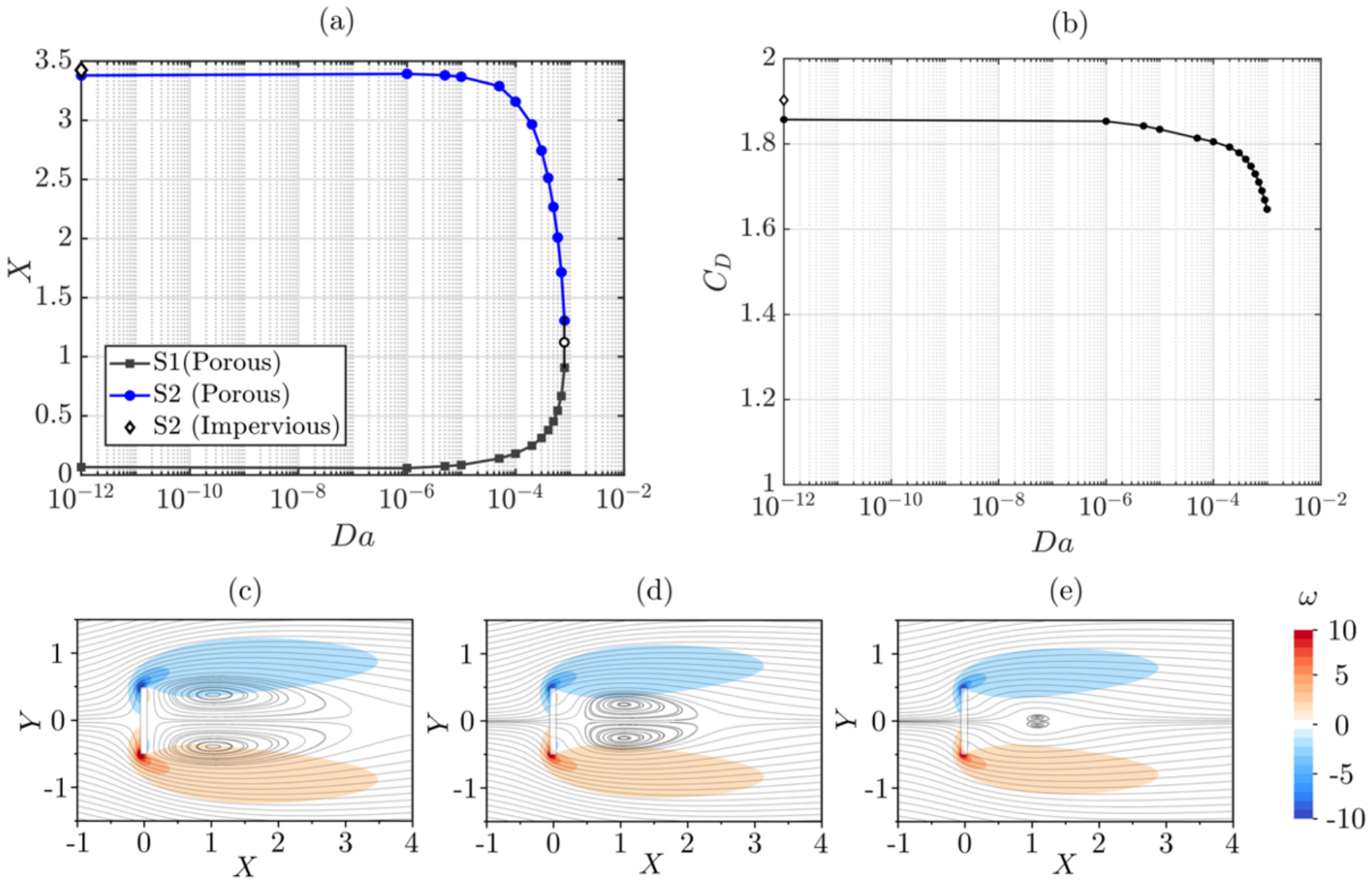

**Fig. 4**: (a) Stream-wise coordinate of saddle points S1 and S2, and (b) drag coefficient versus the Darcy number for a 2D porous plate with $\alpha = 90°$ and $Re = 30$. Empty diamonds at $Da = 10^{-12}$ indicate the values computed for a solid plate at the same $\alpha$ and $Re$. Streamlines and vorticity field, $\omega$, for (c) $Da = 5 \times 10^{-5}$, (d) $Da = 5 \times 10^{-4}$, and (e) $Da = 8 \times 10^{-4}$. The numerical uncertainty for $C_D$ is estimated as 4.65%.

### 3.4 Forces and torque on a porous plate at incidence

Akin to the porous plate normal to the stream, both $C_L$ and $C_D$ decrease with increasing permeability at any $\alpha$ (Fig. 6a and 6b). The lift and the torque vanish at $\alpha = 0°$ and 90°, and their absolute value is maximum for the same critical incidence, $\alpha_{\max}$, which increases with the permeability. Interestingly, at $Da = 5 \times 10^{-5}$, $C_L$ and $|C_M|$ are maximum at $\alpha = 34°$, where the wake is characterised by a single clockwise recirculation region; while at $Da = 10^{-3}$, $C_L$ and $|C_M|$ are maximum near $\alpha = 40°$, where there is no recirculation region. The $C_D$ value, which increases monotonically from $\alpha = 0°$ to 90°, shows a higher reduction with permeability at $\alpha = 90°$ than 0°.

In the body-fixed frame of reference, while $C_x$ decreases with $Da$ (Fig. 6c), $|C_y|$ increases with $Da$ for any $\alpha$ between 20° and 90°; see figs. 6d and 6e. To understand this result, we estimate the force components in the four faces of the porous plate, F1-F4, as defined in the inset of Fig. 6. For each face, we consider the pressure and viscous components of $C_x$ and $C_y$. The plate-normal force coefficient $C_x$ is primarily driven by the suction on F2 and, to a lesser extent, by the pressure on F4 (Fig. 6f). This is akin to a foil at incidence. Both the force contributions decrease with permeability, as expected. In contrast, $C_y$ is primarily driven by the shear on F4 and to a lesser extent, by the suction on F3 (Fig. 6g). As the permeability increases, the overall drag decreases (Fig. 6a), and thus the flow speed and the shear near the intersection of F4 with F3 increase, resulting in an overall increase in $|C_y|$ (Fig. 6e).

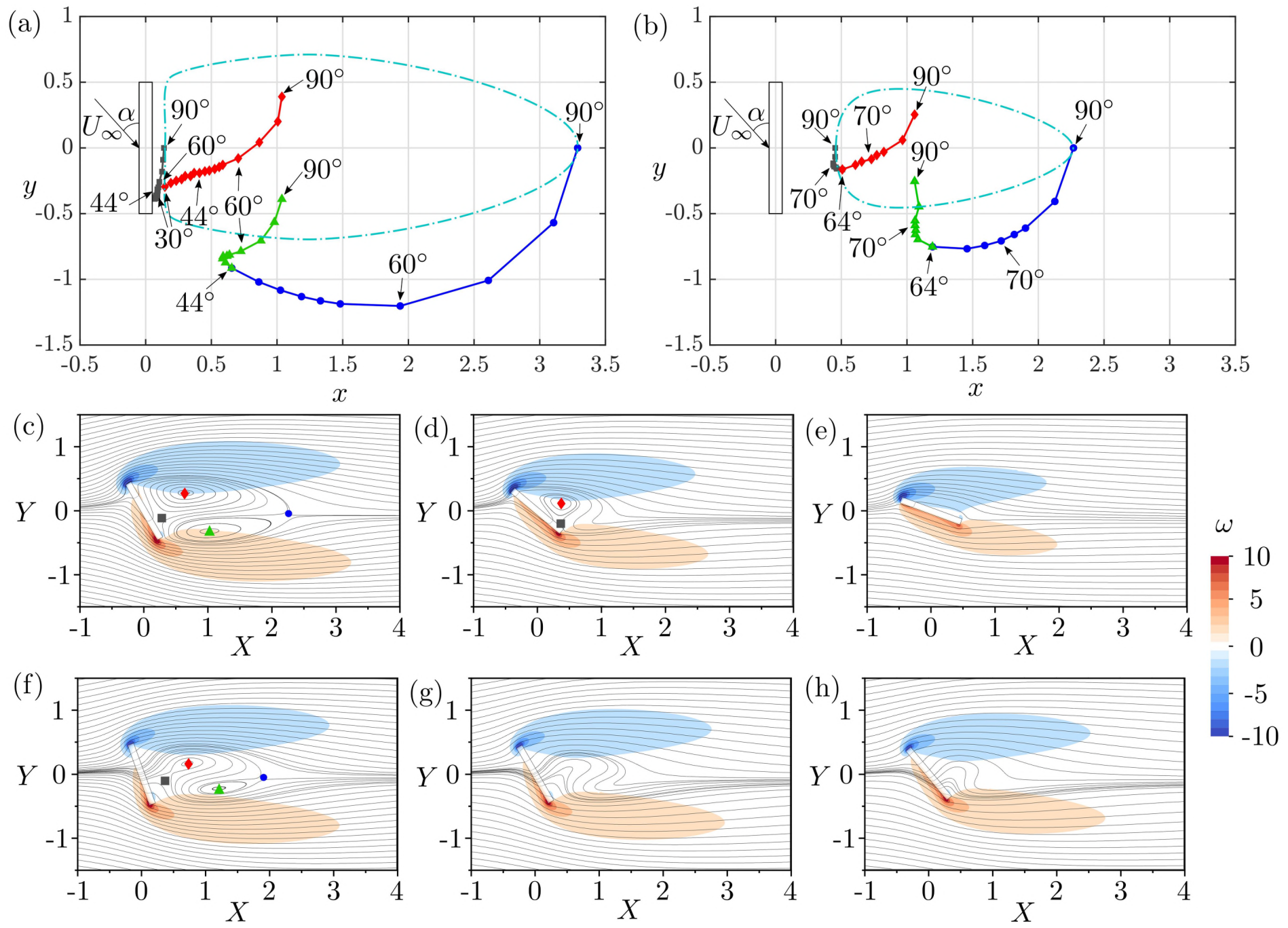

**Fig. 5**: Position of nodes N1 (red diamond), N2 (green triangle), and saddle points S1 (grey square), S2 (blue circle) of a 2D porous plate at $Re = 30$ for (a) $Da = 5 \times 10^{-5}$ and (b) $Da = 5 \times 10^{-4}$, where $\alpha = [90°, 80°, 70°, 60°, 54°, 52°, 50°, 48°, 46°, 44°, 42°, 40°, 38°, 36°, 34°, 32°, 30°]$ in (a), $\alpha = [90°, 80°, 74°, 72°, 70°, 68°, 66°, 64°]$ in (b), and light blue dotted lines denote the closed streamline of the vortex dipole at $\alpha = 90°$. Streamlines and vorticity field for $Da = 5 \times 10^{-5}$ and (c) $\alpha = 60°$, (d) $\alpha = 40°$, and (e) $\alpha = 20°$. Streamlines and vorticity field for $Da = 5 \times 10^{-4}$ and (f) $\alpha = 70°$, (g) $\alpha = 60°$, and (h) $\alpha = 50°$. When present, the positions of N1, N2, S1, and S2 are indicated.

## 3.5 Force calculation from the fluid impulse

We aim to investigate how the vorticity field around the plate varies with the permeability, and how these changes are correlated with the forces. For steady 2D flows, the relationship between the vorticity field and the lift is given by the Kutta-Joukowsky lift formula, which, in nondimensional form, is $L = -\Gamma$, where $\Gamma$ is the nondimensional circulation and $C_L = 2L$. For steady flow conditions, $\Gamma$ should be computed as the integral of vorticity over a region enclosing the plate, such that the net flux of vorticity across the region boundary vanishes [64]. Here, we consider the integral of vorticity within the whole domain.

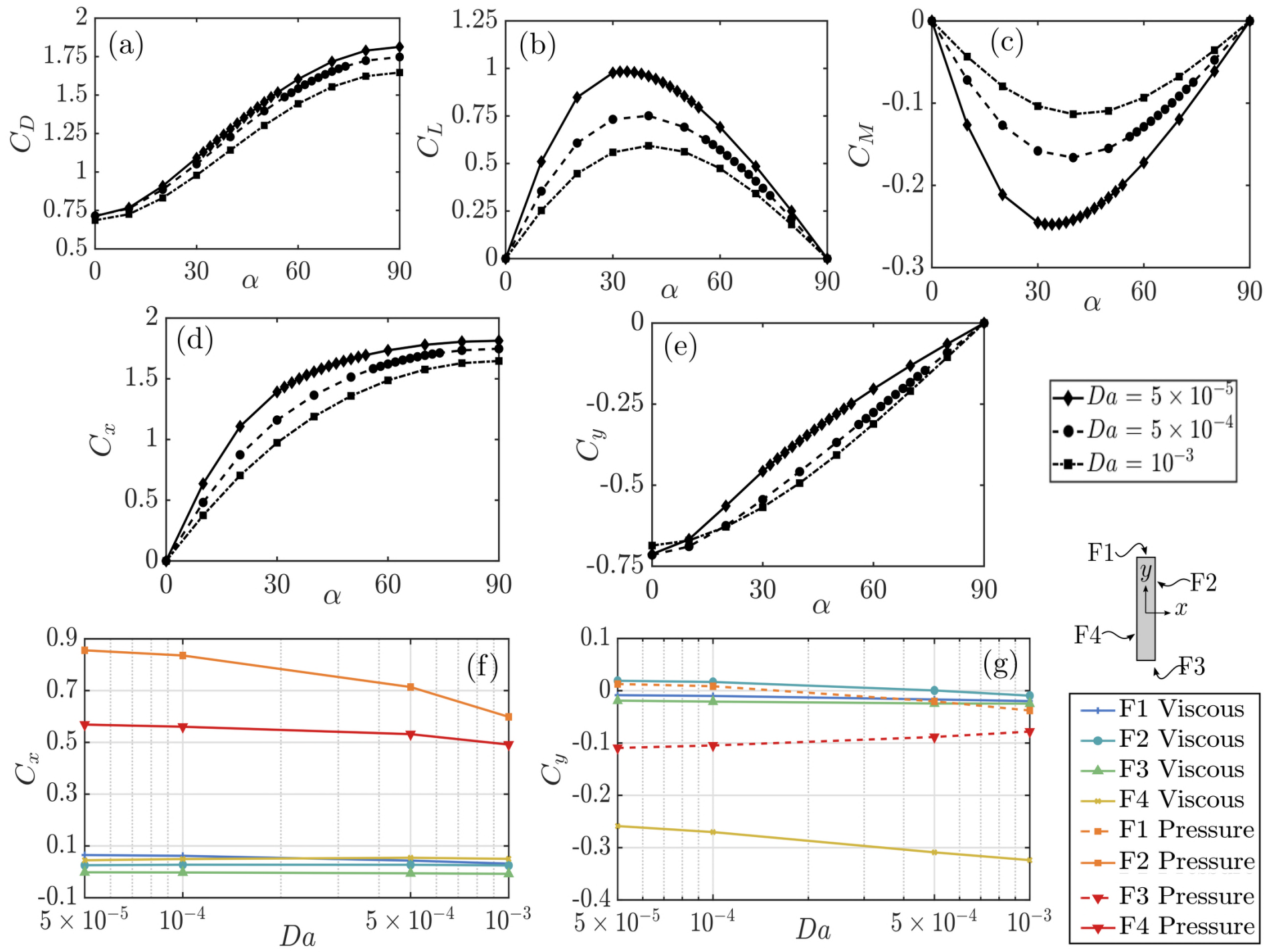

**Fig. 6**: Coefficients of (a) drag, (b) lift (c) torque, (d) plate-normal force (e) plate-wise force for porous plates with three different permeability values versus the flow incidence. (f) Plate-normal and (g) plate-wise pressure and viscous force coefficients for each face (F1, F2, F3 and F4) of a porous plate at $\alpha = 40°$ versus the Darcy number. The numerical uncertainty for the force coefficients is estimated in 4.65%.

For steady 2D flows, the non-dimensional drag $D$ is [64]

$$D = -\int_W Y\omega \, \mathrm{d}W, \tag{4}$$

where the line $W$ orthogonally intersects the far wake, and $C_D = 2D$. Here, we chose $W$ as a stream-normal section of the domain at $X = 70$. The comparison between the force coefficients computed with the impulse theory and with the stress tensor (eq. 3) is shown in Fig. 7.

From the Kutta-Joukowsky lift formula ($L = -\Gamma$), we infer that the difference in the lift of two plates with different $Da$ is proportional to the change in the integral of vorticity in the whole field. To gain insights into the underlying mechanism that leads to a lift change, we set out to investigate which region experiences the greatest change in vorticity due to a change in permeability. We consider the plate at $\alpha = 40°$ and three $Da$ values: a low permeability case with $Da = 5 \times 10^{-5}$, where a single recirculation zone with negative circulation exists (Fig. 5d); a high permeability case with $Da = 5 \times 10^{-4}$, where there are no close recirculation regions

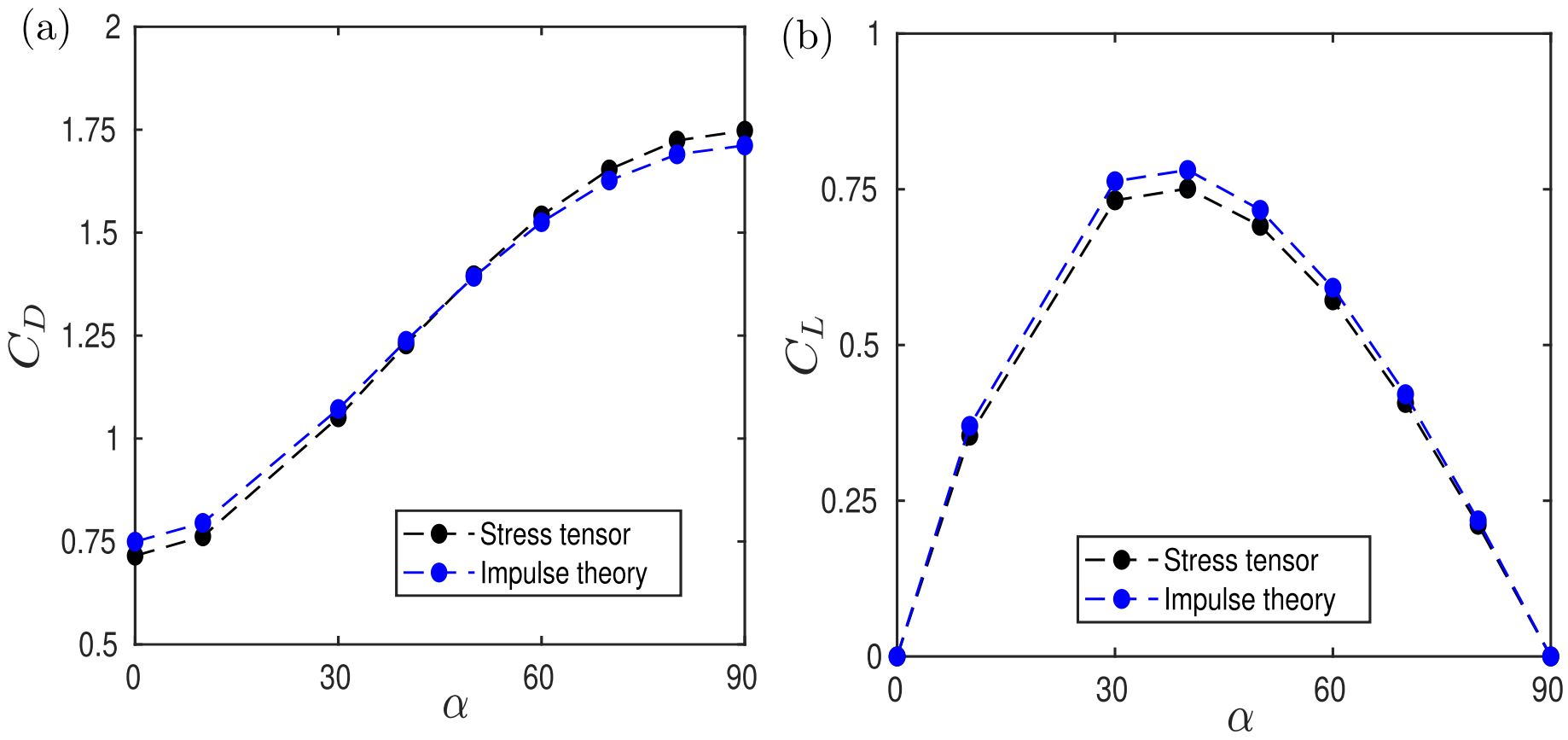

**Fig. 7**: Lift and drag coefficients are computed with the stress tensor and with impulse theory for different incidences at $Da = 5 \times 10^{-4}$. The numerical uncertainty for the force coefficients computed with the stress tensor is estimated in 4.65%.

(Fig. 5g); and the maximum permeability case, investigated in this study, with $Da = 10^{-3}$. Here, the terms 'low-permeability' and 'high-permeability' are used consistently with the previous sections. The vorticity fields of the low, high, and max permeability cases are denoted as $\omega_\mathrm{L}$, $\omega_\mathrm{H}$, and $\omega_\mathrm{M}$, respectively. The lift and drag coefficients for these three cases are shown in table 4.

**Table 4**: Lift and drag coefficients computed with the stress tensor and with impulse theory for $\alpha = 40°$ and three values of the Darcy number. The numerical uncertainty for the force coefficients is estimated in 4.65%.

| | $Da = 5 \times 10^{-5}$ | | $Da = 5 \times 10^{-4}$ | | $Da = 10^{-3}$ | |
|---|---|---|---|---|---|---|
| | $C_L$ | $C_D$ | $C_L$ | $C_D$ | $C_L$ | $C_D$ |
| Conventional | 0.960 | 1.280 | 0.751 | 1.228 | 0.594 | 1.142 |
| Impulse Theory | 1.021 | 1.292 | 0.781 | 1.237 | 0.586 | 1.174 |

The spatial distributions of the differential vorticity fields $\omega_\mathrm{H} - \omega_\mathrm{L}$ and $\omega_\mathrm{M} - \omega_\mathrm{H}$ (figs. 8a and 8b, respectively) show how each fluid region contribute to the change in the lift. Four distinct regions are observed: two associated with the leading and trailing edge shear layers, and two associated with the suction and pressure side of the plate. For example, in Fig. 8, the threshold value of $|\omega_t| = 0.48$ is used to clearly identify the four regions. The vorticity within each region is integrated to compute the contribution to the total change in circulation, and thus, in lift. The percentage values indicate the absolute circulation fraction of each region over the summation of the absolute circulation of all four regions. It is noted that these percentage values are about independent of the threshold value $\omega_t$. For both $\omega_\mathrm{H} - \omega_\mathrm{L}$ and $\omega_\mathrm{M} - \omega_\mathrm{H}$, the integral of the change of vorticity near the leading-edge is substantially higher than the change of vorticity in the other three marked flow regions. Hence, we conclude that the change in the strength of the leading edge shear layer is the main driver for the lift change. Furthermore, its dominant role compared to that of the two shear layers along the chord and that of the trailing edge shear layer increases with $Da$ (e.g. from 48% to 66% in Fig. 8).

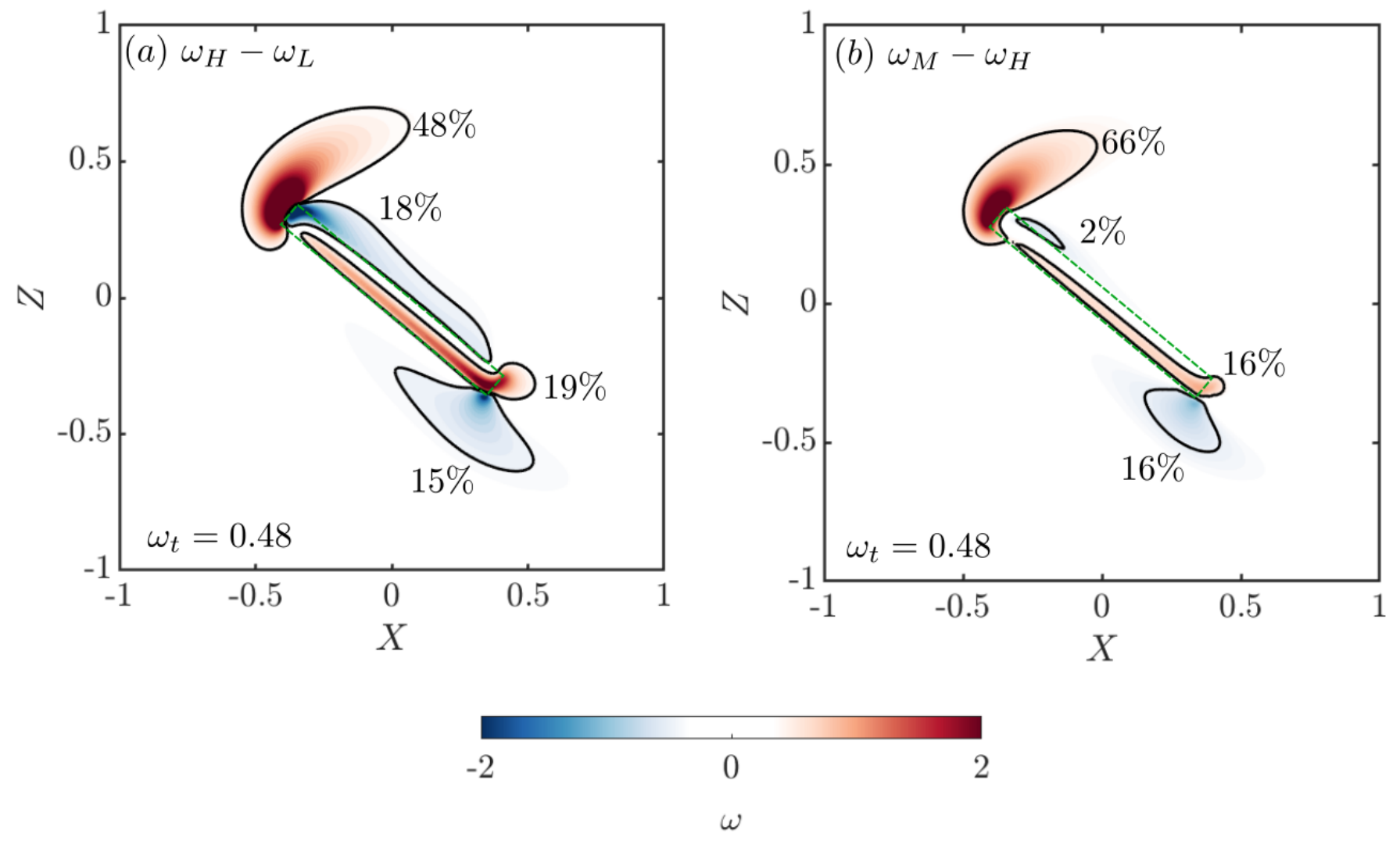

**Fig. 8**: Differential vorticity fields for the porous plate at $\alpha = 40^\circ$: (a) $\omega_H - \omega_L$, and (b) $\omega_M - \omega_H$. The percentage values show the relative circulation of the four regions identified by the isolines of differential vorticity $|\omega| = 0.48$.

To investigate how changes in the vorticity field due to the increased permeability are correlated with the loss of drag, we consider the first moment of vorticity in the stream-wise direction, $Y\omega$, whose integral along the line W is the drag (eq. 4). The leading- and trailing-edge vortex sheet results in two peaks (Fig. 9a), whose amplitude decreases with permeability, while their width is constant. This result suggests that the drop in drag with an increased permeability is primarily due to the weakening of the strength of the vortex sheet and not, for example, by the reduction of the width of the wake. To quantify the relative effect of the leading- and trailing-edge vortex sheet strength on the change of drag coefficient, we consider the difference of the first moment of vorticity in the stream-wise direction for the $\omega_L - \omega_H$ and $\omega_H - \omega_M$ cases; see Fig. 9b. The figure also shows the zeros of the functions $Y|\Delta\omega| = 0$ for $\Delta\omega = \omega_L - \omega_H$ ($A_1$-$D_1$), and for $\Delta\omega = \omega_H - \omega_M$ ($A_2$-$D_2$). By computing the integral along Y between consecutive zeros, we find that for the $\omega_L - \omega_H$ case, the changes in $C_D$ due to the weakening of the leading- and trailing-edge vortex sheet are 0.0251 and 0.0555, respectively; whereas, the corresponding changes in $C_D$ for the $(\omega_H - \omega_M)$ case are 0.0192 and 0.0577, respectively. These results suggest that, with the increased permeability, the weakening of the trailing-edge vortex sheet is more significant than that of the leading-edge vortex sheet on the drag reduction.

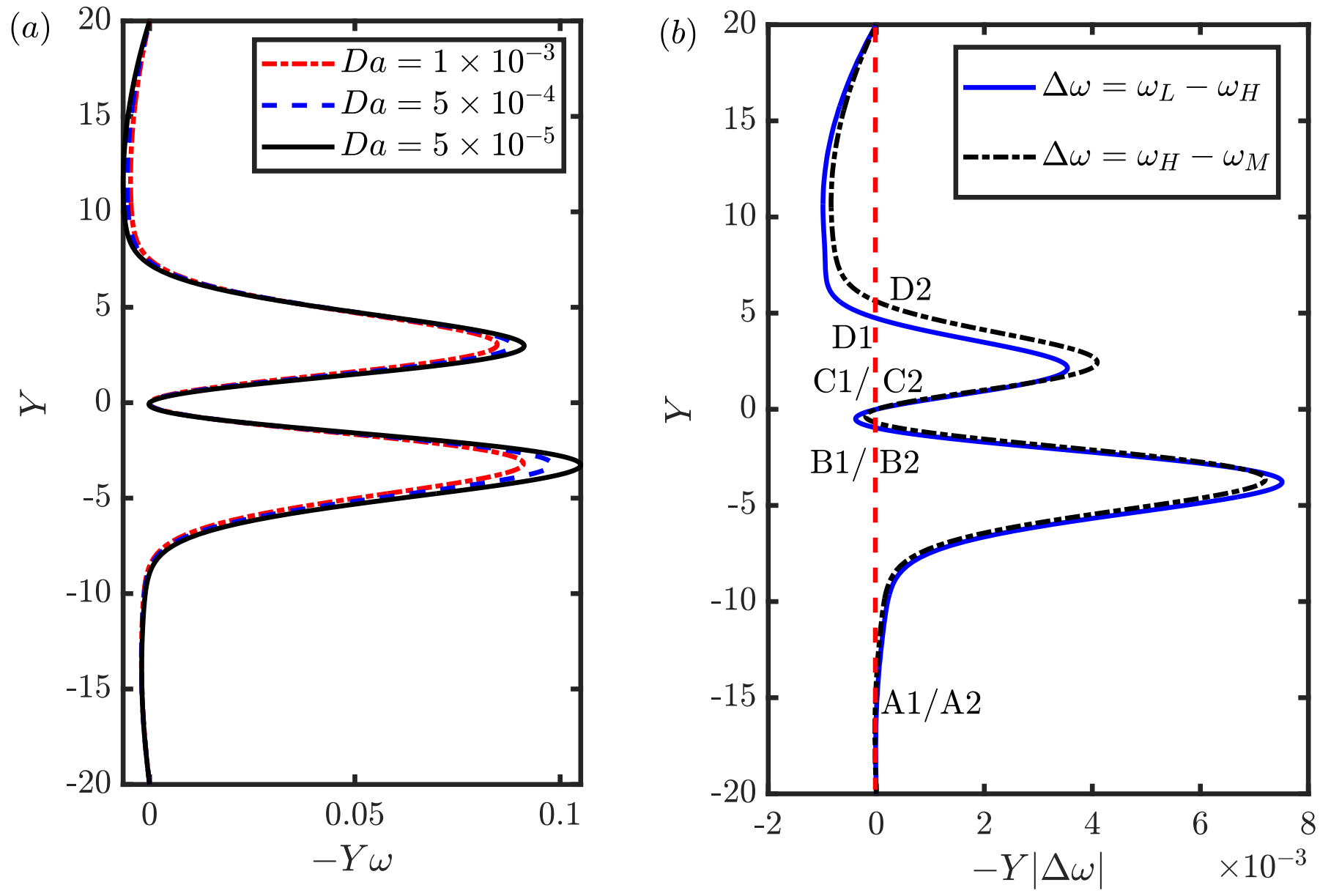

**Fig. 9**: (a) Stream-wise component of the negative first moment of the vorticity field $-Y\omega$ and (b) the difference of the negative first moment of the vorticity field $-Y|\Delta\omega|$ for the $\omega_L - \omega_H$ and $\omega_H - \omega_M$ cases along the $X = 70$ line for $\alpha = 40°$ and $Da = 5 \times 10^{-5}, 5 \times 10^{-4}$ and $10^{-3}$.

# 4 Conclusions

The flow behind a 2D porous plate is investigated for different values of the permeability ($Da$) and flow incidence ($\alpha$). The flow typology is investigated for Reynolds number ($Re$) values between 30 and 90, while a detailed analysis of the forces is undertaken at $Re = 30$, where the wake is found to be steady for any $Da$ and $\alpha$. An incompressible Navier-Stokes solver of the open-source library OpenFOAM has been modified to solve for the Darcy-Brinkman-Forchheimer equation in the porous region.

For a porous plate normal to the stream, below a critical Darcy number, between $8 \times 10^{-4}$ and $9 \times 10^{-4}$, a vortex dipole with two saddles and two nodes is formed in the wake. With increasing $Da$, the separation of the dipole from the plate increases and both the pairs of topological points merge, eventually annihilating the dipole.

With decreasing $\alpha$ from 90° to 0°, the vortex dipole annihilation process is distinct for low and high permeability cases ($Da = 5 \times 10^{-5}$ and $Da = 5 \times 10^{-4}$). For the former, first, the downstream saddle and node merge, forming a single recirculating region with negative circulation. With decreasing $\alpha$ further, the upstream node and saddle merge, annihilating the recirculating region. Conversely, the four topological points for highly permeable plates merge at the same critical incidence. Lift, drag, and torque decrease in magnitude with $Da$, but there exists a range of $\alpha$ and $Da$ where the plate-wise force component increases in magnitude because of the shear force on the pressure side of the plate.

The steady and unsteady transition boundaries are compared for a representative value of $\alpha = 40^\circ$ and the stream-normal condition. It is observed that the wake remains steady for higher *Re* and lower *Da* values when the porous plate is at an incidence as compared to the stream-normal condition.

The analysis of the rate of change of the flow impulse suggests that the effect of an increased permeability is to decrease the lift by weakening the leading-edge shear layer and to decrease the drag by weakening the trailing-edge shear layer.

The main limitation of this study is that 2D equations are solved. While there is evidence in the literature that the wake past impervious cylinders with rectangular cross-sections is two-dimensional at such low Reynolds numbers, this assumption is not proven through the present investigation, where the effect of permeability is introduced. Furthermore, while the steady to unsteady transition of the wake is expected to be a 2D phenomenon, it is possible that this artificial constrain affects the prediction of this transition.

# 5 Acknowledgements

This work was funded by the ERC Consolidator Grant 'Dandidrone' (101001499), and Callum Bruce's scholarship was funded by the EPSRC grant EP/S023801/1.

# Data availability statement

The data that support the findings of this study are available from the first author/corresponding author upon reasonable request.

# Declaration of interests

The authors report no conflict of interest.